\documentclass[twocolumn, showpacs, preprintnumbers, amsmath, amssymb, superscriptaddress, pra]{revtex4}

\usepackage{graphicx}
\usepackage{dcolumn}
\usepackage{bm}
\usepackage{amssymb}

\begin{document}

\title{Multiparameter entangled state engineering using adaptive optics}

\author{Cristian Bonato}
\affiliation{
Department of Electrical and Computer Engineering\\
Boston University, Boston (MA) USA}
\affiliation{
CNR-INFM LUXOR, Department of Information Engineering\\
University of Padova, Padova (Italy)}

\author{David Simon}
\affiliation{
Department of Electrical and Computer Engineering\\
Boston University, Boston (MA) USA}

\author{Paolo Villoresi}
\affiliation{
CNR-INFM LUXOR, Department of Information Engineering\\
University of Padova, Padova (Italy)}

\author{Alexander V. Sergienko}
\affiliation{
Department of Electrical and Computer Engineering\\
Boston University, Boston (MA) USA}

\affiliation{
Department of Physics\\
Boston University, Boston (MA) USA}

\begin{abstract}
We investigate how quantum coincidence interferometry is affected
by  a controllable manipulation of transverse wave-vectors in
type-II parametric down conversion using adaptive optics
techniques. In particular, we discuss the possibility of spatial
walk-off compensation in quantum interferometry and a new effect
of even-order spatial aberration cancellation.
\end{abstract}


\maketitle

\section{Introduction}
Quantum entanglement \cite{schrod35a} is a valuable resource in
many areas of quantum optics and quantum information processing.
One of the most widespread techniques for generating entangled
optical states is spontaneous parametric downconversion (SPDC)
\cite{klyshko67, harris67, giallorenzi68, kleinman68}. SPDC is a
second-order nonlinear optical process in which a pump photon is
split into a pair of new photons with conservation of energy and
momentum. The phase-matching relation establishes conditions to
have efficient energy conversion between the pump and the
downconverted waves, called signal and idler. This condition sets
also a specific relation between the frequency and the emission
angle of down converted radiation. In other words, the quantum
state emitted in the SPDC process cannot be factorized into
separate frequency and wavevector components. This leads to
several interesting effects where the manipulation of a spatial
variable affects the shape of the polarization-temporal
interference pattern. For example, the dependence of
polarization-temporal interference on the selection of collected
wavevectors was studied in detail in \cite{mete04}.

Here we engineer the quantum state in the space of transverse
momentum and we study how this spatial modulation is transferred
to the polarization-spectral domain by means of quantum
interferometry. We will focus on type-II SPDC using birefringent
phase-matching since the correlations between wave-vectors and
spectrum are stronger than employing other phase-matching
conditions.

Our aim is twofold. From one point of view, we study the effect of
spatial modulations on temporal quantum interference. This could
be useful, for example, in quantum optical coherence tomography
(QOCT) \cite{qoct02, qoct03}. When focusing light on a sample with
non-planar surface, the photons will acquire a spatial phase
distribution in the far-field, which may perturb the shape of the
interference dip. Our results will provide a tool to understand
this effect.

From a second point of view, we would like to study and
characterize spatial modulation as a tool for quantum state
engineering. This may find application in the field of quantum
information processing, where it is important to gain a  high
degree of control over the production of quantum entangled states
entangled in one or more degrees of freedom (hyper-entanglement).

We start (Section II) introducing a theoretical model of a type-II
quantum interferometer, comprising the polarization, spectral and
spatial degrees of freedom. A modulation in the wave-vector space
is provided by an adaptive optical setup and equations for the
polarization-temporal interference pattern in the coincidence rate
are derived. In our analysis (Section III) we will highlight and
discuss theoretically two interesting special cases. The first one
is the possibility of restoring high visibility in type-II quantum
interference with large collection apertures. In some situations,
to collect a higher photon flux or a broader photon bandwidth, it
can be useful to enlarge the collection apertures of the optical
system. But, when dealing with type-II SPDC in birefringent
crystals, for large collection apertures the effect of spatial
walkoff introduces distinguishability between the photons, leading
to a reduced visibility of temporal and polarization quantum
interference. We will show that high visibility can be restored
with a linear phase shift along the vertical axis.

The second effect is the spatial counterpart of spectral
dispersion cancellation \cite{franson92, steinberg92a}. In the
limit of large detection apertures, the correlations between the
photons momenta will cancel out the effects of even-order
aberrations, exactly as  in the limit of slow detectors the
frequency correlations cancel out the even-order terms of spectral
dispersion. The experimental demonstration of this effect has been
reported recently \cite{bonatoAberr2008}.

Finally, in Section IV, we introduce a numerical approach for
practical evaluation of the results of the theoretical model,
discussing a few examples. By means of this approximated model, we
will examine under what conditions the even-order aberration
cancellation effect can be observed (Section V).

\section{Theoretical model}

\begin{figure} [ht]
\centering
\includegraphics [width = 9 cm] {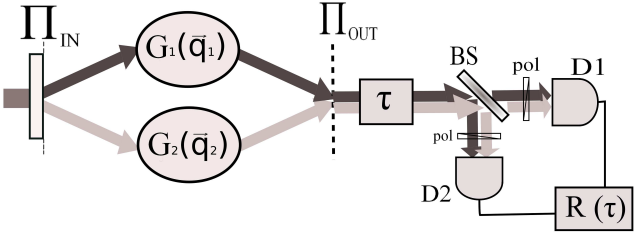}
\caption{\textit{Scheme of the proposed setup.
Horizontally-polarized photons from type-II SPDC are assigned a
phase dependent on the photon transverse momentum $\phi_o (q_o)$,
while the vertically-polarized ones are assigned a phase $\phi_e
(q_e)$. The modulated photons enter a type-II quantum
interferometer, which records the coincidence count rate as a
function of the delay $\tau$ between the photons given by an
appropriate delay-line. }} \label {Fig:schema}
\end{figure}

Consider the scheme in Fig. \ref{Fig:schema}. A laser beam pumps a
$\chi^{(2)}$ nonlinear material phase-matched for type-II
parametric down-conversion, creating a pair of entangled photons.
Each of the generated photons passes through a Fourier transform
system, then enters a modulation system which transforms each
transverse wave-vector for the horizontally polarized photon
according to the transfer function $G_1 (\mathbf{q_1})$ and for
the vertically-polarized photon according to $G_2 (\mathbf{q_2})$.
After being modulated in the $q$-space, the photons enter a
type-II interferometer. A non-polarizing beamsplitter creates
polarization entanglement from the polarization-correlated pair
emitted by the source. The beams at the output ports of the
beamsplitter are directed towards two single photon detectors. Two
polarizers at 45 degrees before the BS restore
indistinguishability in the polarization degree of freedom. An
adjustable delay-line $\tau$ is scanned and the coincidence rate
$R (\tau)$ between the detection events of the two detectors is
recorded. An aperture is placed before the beamsplitter to select
an appropriate collection angle.

\subsection{Notation}
Consider a monochromatic plane wave of complex amplitude $E
(\mathbf{r}) = E_0 e^{-i \mathbf{k} \cdot \mathbf{r}}$ with
$\mathbf{r} = (x, y, z)$. For a given wavelength $\lambda$,
corresponding to a frequency $\Omega$, the wave-vector can be
split in a transverse component $\mathbf{q} = (q_x, q_y)$ and a
longitudinal component $\beta (\mathbf{q}; \Omega)$:
\begin{equation}
\mathbf{k} = [\mathbf{q}, \beta (\mathbf{q}; \Omega)]
\end{equation}
The wave-number is:
\begin{equation}
k (\Omega) = \frac {n (\Omega) \Omega} {c}
\end{equation}
The longitudinal component of the wave-vector is:
\begin{equation}
\beta (\mathbf{q}, \Omega) = \sqrt{k^2 (\Omega) - |\mathbf{q}|^2}
\end{equation}
Therefore the electric field at the position $\mathbf{r}$ and time $t$ can be written as:
\begin{equation}
E (\mathbf{r}; t) = \int d\mathbf{q} \int d\Omega \tilde{E} (\mathbf{q}, \Omega) e^{-i \mathbf{q} \cdot \mathbf{\rho}} e^{-i \beta(\mathbf{q}; \Omega)z} e^{i \Omega t}
\end{equation}
where $\mathbf{\rho} = (x, y)$.

In paraxial approximation $|\mathbf{q}|^2 \ll k^2 (\Omega)$, so that:
\begin{equation}
\beta (\mathbf{q}, \Omega) \approx k (\Omega) - \frac{|\mathbf{q}|^2}{2 k(\Omega)}
\end{equation}
For a quasi-monochromatic wave-packet centered around the frequency $\Omega_0$ one can write $\Omega = \Omega_0 + \nu$, with $\nu \ll \Omega_0$ and this expression can be approximated by:
\begin{equation}
\beta (\mathbf{q}, \Omega) \approx k_0 + \frac {\nu}{u_0} - \frac{|\mathbf{q}|^2}{2 k_0}
\end{equation}
where $k_0 = k (\Omega_0)$ and $u_0 = \left( \frac{d k(\Omega)}{d\Omega} \vert_{\Omega = \Omega_0}\right)^{-1}$ is the group velocity for the propagation of the wave-packet through the material.

\subsection{State Generation}
Using first-order time-dependent perturbation theory the two-photon state at the output of the nonlinear cystal can be calculated as:
\begin{equation}
\left | \psi \right \rangle \sim -\frac{i}{\hbar} \int dt H_I (t) \left | 0 \right \rangle
\end{equation}
where the interaction Hamiltonian is:
\begin {equation}
H_I (t) = \frac{1}{V} \int d\mathbf{r} \chi^{(2)} (\mathbf{r})
E_p^{(+)} (\mathbf{r}, t) E_s^{(-)} (\mathbf{r}, t) E_i^{(-)}
(\mathbf{r}, t)
\end {equation}
The strong, undepleted pump beam can be treated classically.
Assuming a monochromatic plane-wave propagating along the z
direction:
\begin{equation}
E_p (\mathbf{r}, t) = E_p e^{i(k_p z - \omega_p t)}
\end{equation}
The signal and idler photons are described by the following quantum field operators:
\begin{equation}
\hat{E}_i^{(-)} (\mathbf{r}, t) = \int d\mathbf{q}_i \int
d\omega_i e^{i[\beta(\mathbf{q}_i, \omega_i) z + \mathbf{q}_i
\cdot \mathbf{\rho} - \omega_i t]} \hat{a} (\mathbf{q}_i,
\omega_i)
\end{equation}

The biphoton quantum state at the output plane of the nonlinear crystal is \cite{rubin96}:
\begin{equation}
\label {psi_gen} \left | \psi \right \rangle = \int d\mathbf{q}
\int d\nu \tilde{\Phi} (\mathbf{q}, \nu) \hat{a}_o^{\dagger}
(\mathbf{q}, \Omega_0 + \nu) \hat{a}_e^{\dagger} (\mathbf{-q},
\Omega_0 - \nu) \left | 0 \right \rangle
\end{equation}
Two photons are emitted from the nonlinear crystal, one
horizontally-polarized (ordinary photon) and the other
vertically-polarized (extraordinary photon), with anticorrelated
frequencies and emission directions.

In the case of a single bulk crystal of thickness L and constant
nonlinearity $\chi_o$, the probability amplitude for having the
signal photon in the mode $(\mathbf{q}, \Omega_0 + \nu)$ and the
idler in the mode $(-\mathbf{q}, \Omega_0 - \nu)$:

\begin{equation}
\label {phi_bulk}
\tilde{\Phi} (\mathbf{q}, \nu) = \mbox {sinc} \left[ \frac {L \Delta(\mathbf{q}, \nu)}{2}\right] e^{i  \frac {\Delta(\mathbf{q}, \nu)L}{2}}
\end{equation}
For type-II collinear degenerate phase-matching, the phase-mismatch function $\Delta (\mathbf{q}, \nu)$ can be approximated to be:
\begin{equation}
\label {Delta_k}
\Delta(\mathbf{q}, \nu) = -\nu D +  M \mathbf {\hat{e}_2}\cdot \mathbf{q} + \frac{2|\mathbf{q}|^2}{k_p}
\end{equation}
where D is the difference between the inverse of the group velocities of the ordinary and extraordinary photons inside the birefringent crystal and the quadratic term in $\mathbf{q}$ is due to diffraction in paraxial approximation. The last term is the first-order approximation for the spatial walk-off.

\subsection{Propagation}
Consider a photon described by the operator $\hat{a}_j
(\mathbf{q}, \Omega)$ (polarization $j = e, o$, frequency
$\Omega$, and transverse momentum $\mathbf{q}$). Its propagation
through an optical system to a point $\mathbf{x}_k$ on the output
plane is described by the optical transfer function $H_j
(\mathbf{x}_k, \mathbf{q}; \Omega)$. In our setup, the field at
the detector will be a superposition of contributions from the
ordinary and extraordinary photons. The quantized electric fields
at the detector planes are:
\begin{equation}
\label {fields}
\begin {split}
\hat{E}_A^{(+)} (\mathbf x_A, t_A) & = \int d\mathbf{q} \int d\omega e^{i \omega t_A} [H_{e} (\mathbf{x_A}, \mathbf {q}; \omega) \hat{a}_e (\mathbf{q}, \omega) \\
& \qquad + H_{o} (\mathbf{x_A}, \mathbf {q}; \omega) \hat{a}_o (\mathbf{q}, \omega)]\\
\hat{E}_B^{(+)} (\mathbf x_B, t_B) & = \int d\mathbf{q} \int d\omega e^{i \omega t_B} [H_{e} (\mathbf{x_B}, \mathbf {q}; \omega) \hat{a}_e (\mathbf{q}, \omega) +\\
 & \qquad H_{o} (\mathbf{x_B}, \mathbf {q}; \omega) \hat{a}_o (\mathbf{q}, \omega)]
\end {split}
\end{equation}
The probability amplitude to detect a photon pair at the detector
planes, with space-time coordinates $(\mathbf{x_A}, t_A)$ and
$(\mathbf{x_B}, t_B)$, is:
\begin{equation}
\label {prob_ampl}
A (\mathbf{x_A}, \mathbf{x_B}; t_A, t_B) = \left \langle 0 \right |\hat{E}_A^{(+)} (\mathbf x_A, t_A) \hat{E}_B^{(+)} (\mathbf x_B, t_B) \left | \psi \right \rangle
\end{equation}

For the biphoton probability amplitude we get:

\begin{equation}
\label {Eq:prob_ampl2}
\begin {split}
& A (\mathbf{x_A}, \mathbf{x_B}; t_A, t_B) = \int d\mathbf{q_o} d\mathbf{q_e} d\omega_o d\omega_e \Phi (\mathbf{q_o}, \mathbf{q_e}; \omega_o, \omega_e) \\
& \qquad \Bigl[ H_{e} (\mathbf{x_A}, \mathbf{q_e}; \omega_e) H_{o} (\mathbf{x_B}, \mathbf{q_o}; \omega_o) e^{-i(\omega_e t_A + \omega_o t_B)} + \\
& \qquad H_{o} (\mathbf{x_A}, \mathbf{q_o}; \omega_o) H_{e}
(\mathbf{x_B}, \mathbf{q_e}; \omega_e) e^{-i(\omega_o t_A +
\omega_e t_B)} \Bigr]
\end {split}
\end{equation}

This probability amplitude represents the superposition of two possible events leading to a coincidence count in the detectors:
\begin{enumerate}
 \item the V-polarized photon with momentum $\mathbf{q_e}$ and frequency $\omega_e$ going through the lower branch to arrive at position $x_A$ in detector
 A, while
 the H-polarized photon with momentum $\mathbf{q_o}$ and frequency $\omega_o$ goes through the upper branch to arrive at position $x_B$ in detector
 B.
 \item the V-polarized photon with momentum $\mathbf{q_e}$ and frequency $\omega_e$ going through the lower branch to arrive at position $x_B$ in
detector B, while the H-polarized photon with momentum
$\mathbf{q_o}$ and frequency $\omega_o$ goes through the upper
branch to arrive at position $x_A$ in detector A.
\end{enumerate}
Since the superposition is coherent, there are quantum interference effects between the two probabilities amplitudes.

\subsubsection{State engineering section}
In the state engineering section, each of the two branches
consists of a pair of achromatic Fourier-transform systems coupled
by a spatial light modulator or a deformable mirror. Each
Fourier-transform system consists of a single lens of focal length
$f$, separated from the optical elements before and after it by a
distance $f$. The first Fourier system maps each incident
transverse wave-vector $\mathbf{q}$ on the plane $\Pi_{IN}$ to a
point $\mathbf{x} (\mathbf{q})$ on the Fourier plane $\Pi_{F}$:
\begin{equation}
 \mathbf{x} (\mathbf{q}) = \frac{f}{k_0} \mathbf{q} \qquad k_0 =\frac{\Omega_0}{c}
\end{equation}
where f is the focal length of the Fourier-transform system. Since
we assume the system is achromatic for a certain bandwidth around
a central frequency $\Omega_0$, the position $\mathbf{x}
(\mathbf{q})$ depends only on $\mathbf{q}$ and not on $\omega$.

The spatial modulator assigns a different amplitude and phase to
the light incident on each point, as described by the function $G
(\mathbf{x}) =  t (\mathbf{x}) e^{i \varphi (\mathbf{x})}$. Each
point is then mapped back to a wave-vector on the plane
$\Pi_{OUT}$ by the second achromatic Fourier-transform system.

Using the formalism of Fourier optics \cite{goodmanFO}, the transfer function between the planes $\Pi_{IN}$ and $\Pi_{OUT}$ can be calculated to be:
\begin{equation}
\label {h1}
h_1 (\mathbf{x_1}, \mathbf {x_3}) = \int d\mathbf{x} \quad G(\mathbf{x}) \quad e^{-i \frac{k_0}{f} \mathbf{x} \cdot (\mathbf{x_1} + \mathbf{x_3})}
\end{equation}
The corresponding momentum transfer function is:
\begin{equation}
\label {H1}
H_1 (\mathbf{q_1}, \mathbf {q_3}) = G \left[ \frac{f}{k_0}\mathbf{q_1}\right]  \delta (\mathbf{q_1} - \mathbf{q_3})
\end{equation}

\subsubsection{Interferometer}
After the plane $\Pi_{OUT}$ the two photons enter a type-II
quantum interferometer. Each propagates in free space to a
birefringent delay-line and a detection aperture $p (\mathbf{x})$
to be finally focused to the detection planes by means of lenses
of focal length $f_0$. Following the derivation in \cite{mete04}
the transfer function is:

\begin {eqnarray}
H_2(\mathbf{x_i}, \mathbf{q}; \omega) &=& \int
h(\mathbf{x_1},\mathbf{x_i};\omega ) e^{i \mathbf{q}\cdot
\mathbf{x_1}}d\mathbf{x_1} \nonumber \\& =& e^{i(\omega/c) (d_1 +
d_2 + f_0)} \mbox{exp}
\left[ -i \frac{\omega |\mathbf{x_i}|^2}{2cf_0} \left( \frac{d_2}{f_0} - 1\right) \right] \nonumber\\
& & \qquad \cdot \; e^{-i(c d_1/2 \omega) |\mathbf{q}|^2}
\tilde{P} \left( \frac {\omega}{cf_0} \mathbf{x_i} - \mathbf{q}
\right) \label {H2}
\end {eqnarray}
where $\tilde {P} (\mathbf{q})$ is the Fourier transform of $|p
(\mathbf{x})|^2$.

A combination of the two different stages is described by the transfer function:
\begin{equation}
\label{Eq:acca_j} H_{\alpha} (\mathbf{x_j}, \mathbf{q_{\alpha}};
\omega_{\alpha}) = G_{\alpha} \left[
\frac{f}{k_0}\mathbf{q_{\alpha}}\right] H_2(\mathbf{x_j},
\mathbf{q_{\alpha}}; \omega) ,
\end{equation}
where the two functions $G_1 (\mathbf{q})$ and $G_2 (\mathbf{q})$
are the momentum transfer function which describe the modulation
imparted respectively on the ordinary and the extraordinary
photon.

\subsection{Detection}
Since the single-photon detectors used in quantum optics
experiments are slow with respect to the temporal coherence of the
photons and their area is larger than the spot into which the
photons are focused by the collection lens, we integrate over the
spatial and temporal coordinates. Therefore the coincidence
count-rate expressed in terms of the biphoton probability
amplitude is:
\begin{equation}
\label {R_tau1}
R (\tau) = \int d\mathbf{x_A} \int d\mathbf{x_B} \int dt_A \int dt_B |A (\mathbf{x_A}, \mathbf {x_B}; t_A, t_B)|^2
\end{equation}

Following the derivation described in Appendix A, one gets:
\begin{equation}
\label {Eq:Rtau}
R (\tau) = R_0 - \Lambda \left( 1 - \frac {2\tau}{DL}\right) W_G (\tau)
\end{equation}
where $\Lambda (x)$ is the triangular function:
\begin {equation}
\Lambda (x) =
\bigg \{
\begin{array}{l}
1 - |x|, \qquad |x| \leq 1 \\
0, \qquad \qquad |x| > 1
\end{array}
\end{equation}

Therefore, the coincidence count rate $R(\tau)$ is given by the
summation of a background level $R_0$ and an interference pattern
given by the triangular dip, $\Lambda \left( 1 - \frac
{2\tau}{DL}\right)$, that one gets when working with narrow
apertures, modulated by the function $W_M (\tau)$ which depends on
the details of the adaptive optical system.

The expressions for $R_0$ and $W_M (\tau)$ are:
\begin{widetext}

\begin {equation}
 \label {R0}
 \begin {split}
 R_0 &= \int d\mathbf{q} \int d\mathbf {q'} \mbox {sinc}
 [ML \mathbf {\hat{e}_2}\cdot (\mathbf{q} - \mathbf{q'})] G_1^* \left(
 \frac{f}{k_0}\mathbf{q}\right)  G_1\left( \frac{f}{k_0}\mathbf{q'}\right) G_2^* \left( -\frac{f}{k_0}\mathbf{q}\right)  G_2\left( -\frac{f}{k_0}\mathbf{q'}\right) \\
 & \qquad \qquad \qquad \qquad e^{-i \frac {ML}{2}\mathbf {\hat{e}_2}\cdot (\mathbf{q} - \mathbf{q'})}
 \quad e^{i \frac {2 d_1}{k_p} [|\mathbf{q}|^2 - |\mathbf{q'}|^2]} \tilde{P}_A (\mathbf{q} - \mathbf{q'}) \tilde{P}_B (-\mathbf{q} + \mathbf{q'})
 \end {split}
 \end {equation}

and

\begin {equation}
\label {Eq:W}
 \begin {split}
 W_M (\tau) &= \int d\mathbf{q} \int d\mathbf {q'} \mbox {sinc} \left[ ML \mathbf {\hat{e}_2}\cdot (\mathbf{q} + \mathbf{q'})\Lambda \left( 1 - \frac {2\tau}{DL}\right) \right] G_1^* \left( \frac{f}{k_0}\mathbf{q}\right)  G_1 \left( \frac{f}{k_0}\mathbf{q'}\right) \\
 &\qquad \qquad \qquad G_2^* \left( -\frac{f}{k_0}\mathbf{q}\right)  G_2 \left(-\frac{f}{k_0}\mathbf{q'}\right) e^{-i \frac {M}{D} \tau \mathbf {\hat{e}_2}\cdot (\mathbf{q} - \mathbf{q'})} e^{i \frac {2 d_1}{k_p} [|\mathbf{q}|^2 - |\mathbf{q'}|^2]} \tilde{P}_A \left[ \mathbf{q} + \mathbf{q'}\right]  \tilde{P}_A \left[ - \left( \mathbf{q} + \mathbf{q'}\right) \right]
 \end {split}
 \end {equation}

\end{widetext}

In the following we will assume there is spatial modulation only
on one of the photons, therefore having $G_2 (\mathbf{q}) \equiv
1$.

 \section{Particular cases}
Let's examine Eq. \ref{Eq:Rtau} in a few simple cases.

First we will consider the case when no spatial modulation is
assigned to the photons and Eq. \ref{Eq:Rtau} will reduce to the
results already described in the literature for quantum
interferometry with multiparametric entangled states from type-II
downconversion \cite{mete04}. Then we will examine the effect of a
linear phase, describing its implications for the compensation of
the spatial walk-off between the two photons. Finally we will
describe what happens in the approximation of sufficiently large
detection apertures, introducing the effect of even-order
aberration cancellation.

 \subsection{No phase modulation}
 Applying no phase modulation, our equations reduce to the ones derived in \cite{mete04}. Particularly we find:
 \begin {equation}
 \label {noMod_R0}
 R_0 = \tilde {P}_A (\mathbf{0}) \tilde {P}_B (\mathbf{0})
 \end {equation}
 and
 \begin {equation}
 \label {noMod_Rt}
 \begin {split}
 W_G (\tau) &= \mbox {Sinc} \left[ \frac{M^2Lk_p}{2d_1D} \tau \Lambda \left( 1 - \frac{2\tau}{DL}\right) \right] \\
 & \qquad \qquad \tilde {P}_A \left[ \frac{M k_p}{2 d_1 D} \tau \mathbf {\hat{e}_2} \right] \tilde {P}_B \left[ -\frac{M k_p}{2 d_1 D} \tau \mathbf {\hat{e}_2} \right]
 \end {split}
 \end {equation}
The shape of the interference pattern is essentially given by the
product of the triangular function by the Fourier-transform of the
aperture function, centered at $\tau = 0$. For physically relevant
parameters the sinc function is almost flat in the region where
the triangular function is not zero.

 \begin{figure} [ht]
\centering
\includegraphics [width = 8 cm] {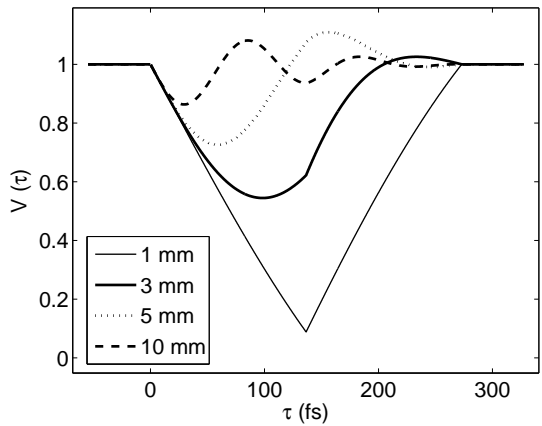}
\caption{\textit{On the right side we can see the interference patterns with three different detector aperture sizes: the corresponding aperture functions are shown on the left side.}}
\label {Fig:noMod}
\end{figure}

To get an analytic result one may assume Gaussian detection
apertures of radius $R_G$ centered along the system's optical
axis:
 \begin {equation}
 \label {gauss_apert}
 p (\mathbf{x}) = e^{-\frac{|\mathbf{x}|^2}{2R_G^2}}
 \end {equation}
In this case the solution is quite simple:
\begin {equation}
\label {noMod_Vt}
R (\tau) = R_0 \left[ 1 - \Lambda \left( 1 - \frac {2\tau}{DL}\right) e^{-\frac{\tau^2}{2\tau_1^2}} \right]
\end {equation}
with:
 \begin {equation}
 \label {noMod_tau1}
 \tau_1 = \frac{2d_1D}{k_pMR_G}
 \end{equation}

Typically, sharp circular apertures are used in experiments. In
this case, the function  $\tilde {P} [\mathbf{q}]$ is described in
terms of the Bessel function $J_1 (x)$. For a circular aperture of
radius $R$:
\begin{equation}
\tilde {P} [\mathbf{q}] = \frac{J_1 (2R|\mathbf{q}|)}{R|\mathbf{q}|}
\end{equation}
However the Gaussian approximation is still a good one if the width $R_G$ of the Gaussian is taken to roughly fit the Bessel function (of width $R$): in our case we take $R_G = R/(2\sqrt{2})$.\\
 Therefore Eq. (\ref{noMod_Vt}) is still approximately valid in the case of sharp circular apertures, just taking:
 \begin {equation}
 \label {noMod_tau1b}
 \tau_1 = \frac{4\sqrt{2}d_1D}{k_pMR_B}
 \end{equation}

Mathematically, in Eq. (\ref{noMod_Vt}) the interference pattern
is given by the multiplication of a triangular function centered
at $\tau = DL/2$ by a Gaussian function centered at $\tau = 0$.
The width of the Gaussian function $\tau_1$ decreases with
increasing radius of the aperture $R_B$. Therefore, in the small
aperture approximation, the width of the Gaussian is so large that
it is approximately constant between $\tau = 0$ and $\tau = DL/2$,
giving the typical triangular dip found in quantum interference
experiments. On the other hand, increasing the aperture size, the
width of the Gaussian function decreases, reducing the dip
visibility (see Fig. \ref {Fig:noMod} ). Physically, this can be
explained by the fact that increasing the aperture size we let
more wave-vectors into the system, and so the spatial walk-off in
type-II interferometry introduces distinguishability, reducing the
interference visibility. Enlarging the aperture sizes is often
useful in practice, for example to get a higher photon flux.
Moreover, since in the SPDC process different frequency bands are
emitted at different angles it may be necessary to open the
detection aperture in applications where a broader bandwidth is
needed. This is clearly a problem when using type-II phase
matching in birefringent crystals, since the visibility of
temporal and polarization interference gets drastically reduced.

\subsection{Linear phase shift}
Suppose now we introduce a linear phase function with the spatial
light modulator, along the directon $\mathbf{s_1}$:
 \begin {equation}
 \label {linMod_def}
 \varphi (\mathbf{x}) = \mathbf{s_1} \cdot \mathbf{x}
 \end {equation}
 we get:

\begin{equation}
\label {linMod_Vt}
 \begin {split}
 W_M (\tau) &=  \mbox {Sinc} \left[ \frac{M^2Lk_p}{2d_1D} \tau \Lambda \left( 1 - \frac{2\tau}{DL}\right) \right] \\
 & \tilde {P}_A \left[ \frac{M k_p}{2 d_1 D} \tau \mathbf {\hat{e}_2} + 2f \mathbf{s_1} \right] \tilde {P}_B \left[ -\frac{M k_p}{2 d_1 D} \tau \mathbf {\hat{e}_2} - 2f \mathbf{s_1} \right]
 \end {split}
\end{equation}

 \begin{figure} [ht]
\centering
\includegraphics [width = 8 cm] {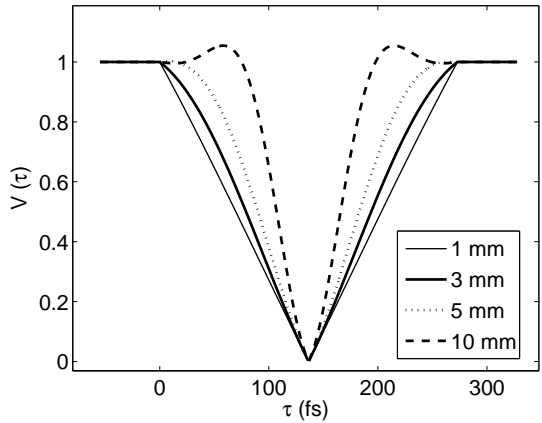}
\caption{\textit{Quantum interference pattern for different
detector aperture sizes, introducing a linear modulation of the
deformable mirror, in order to restore the indistinguishability
between the photons, decreased by the spatial walkoff in the
generation process.}} \label{Fig:linMod1}
\end{figure}

If we compare Eq. (\ref{linMod_Vt}) with Eq. (\ref{noMod_Vt}) we
can see that the structure is the same; we again have a triangular
function centered in $\tau = DL/2$, along with two aperture
functions. But this time, instead of being centered at $\tau = 0$,
the aperture functions can be shifted at will along the $\tau$
axis. Suppose we now apply a tilt along the y axis ($s_{1x} = 0$);
the modulation function is then shifted to:
\begin{equation}
\tau_{center} = \frac{fD}{k_0 M} s_{1y}
\end{equation}
To get the highest possible visibility, the center of the
modulation function must be matched to the center of the
triangular dip:
\begin{equation}
\tau_{center} = \frac{DL}{2}
\end{equation}
so that:
\begin{equation}
s_{1y} = \frac{k_0 ML}{2f}
\end{equation}

In the case of a reflective system, in which the phase modulation
is implemented by means of a deformable mirror, tilted by an angle
$\theta$:
\begin{equation}
 \varphi (\mathbf{x}) =  2 k_0 \tan \theta y = s_{1y} y
\end{equation}
Therefore, the amount of tilt necessary to restore high-visibility is:
\begin{equation}
\tan \theta = \frac{ML}{4f}
\end{equation}
In the case of a 1.5 mm crystal, with M = 0.0723 (pump at 405 nm,
SPDC at 810 nm) and lenses of focal length 20 cm in the 4-f
system, we get:
\begin{equation}
\theta = 0.14 \; mrad
\end{equation}

\subsection{Large aperture approximation}
 If the detection apertures are large enough for the $\tilde{P}_i$ function to be successfully approximated by a delta-function we get:
 \begin {equation}
 \label{LA_Vt}
W_M (\tau) = \int d\mathbf{q}  G_1^* \left( \frac{f}{k_0}\mathbf{q}\right)  G_1\left( -\frac{f}{k_0}\mathbf{q}\right) e^{i \frac{2Mk}{fD}\tau \mathbf {\hat{e}_2}\cdot\mathbf{q}}
\end{equation}

Suppose that the spatial modulator is a circular aperture with radius r, with unit transmission and phase modulation described by the function $\varphi (\mathbf{x})$:

\begin {equation}
G_1 (\mathbf{x}) =
\bigg \{
\begin{array}{cc}
0  & \mbox{if $|\mathbf{x}|>r$} \\
e^{i\varphi(\mathbf{x})} &  \mbox{if $|\mathbf{x}|<r$}
\end{array}
\end {equation}

 In this case the function $\varphi (\mathbf{x})$ can be expanded on a set of polynomials which are orthogonal on the unit circle, like the Zernicke polynomials:
\begin{equation}
 \label {LA_zernike}
 \begin{split}
 \varphi (\mathbf{q}) &= \sum_{n} \sum_{m} \varphi_{nm} R_n^m (\rho)\cos (m\theta) \\
& \qquad \qquad (m = -n, -n+2, -n+4,..., n)
 \end{split}
\end{equation}
 where $\mathbf{q} = (\rho \cos\theta, \rho \sin\theta)$.
 To calculate $\varphi (-\mathbf{q})$ we note that $-\mathbf{q} = [\rho \cos (\theta+\pi), \rho \sin (\theta + \pi)]$, so:
 \begin{equation}
 \label {LA_PHIminusQ}
 \varphi (-\mathbf{q}) = \sum_{n} \sum_{m} R_n^m (\rho)\cos [m(\theta+\pi)]
 \end{equation}
 If m is even then $\cos [m(\theta+\pi)] = \cos (m\theta)$, otherwise if m is odd $\cos [m(\theta+\pi)] = -\cos  m\theta)$. Therefore:
 \begin{equation}
 \label {LA_dispCanc}
 \varphi (\mathbf{q}) - \varphi (-\mathbf{q}) = 2 \sum_n \sum_{\mbox{m odd}} \varphi_{nm} R_n^m (\rho) \cos (m\theta)
 \end{equation}

 So, only the Zernike polynomials with m odd contribute to the shape of the interference pattern. This effect is the spatial counterpart of the dispersion cancellation effect, in which only the odd-order terms in the Taylor expansion of the spectral phase survive.
The experimental demonstration of this effect has been recently reported in \cite{bonatoAberr2008}

\begin{figure} [ht]
\centering
\includegraphics [width = 7 cm] {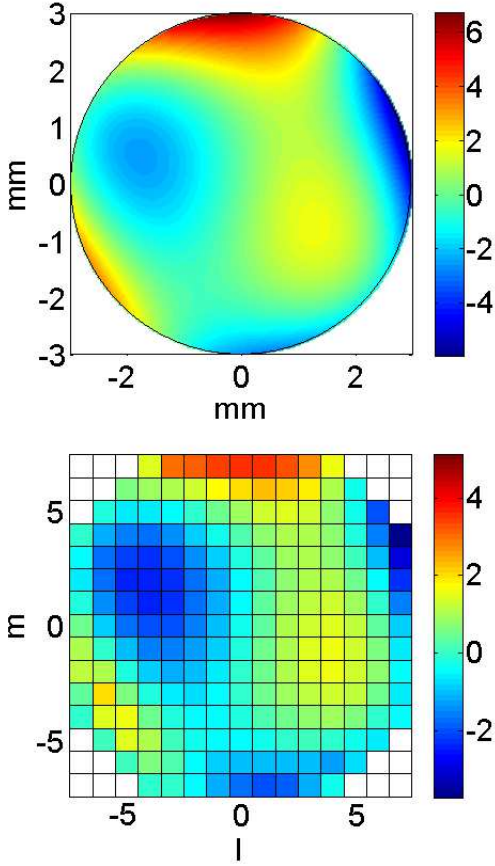}
\caption{Example of the numerical approach adopted to evaluate Eq.
(\ref{R0}) and Eq. (\ref{Eq:W}). The spatial modulation surface is
discretized in sufficiently small squares over which the phase is
averaged. }
\end{figure}

\section {Numerical solutions}
Numerically solving for the quantities in Eq. (\ref{R0}) and Eq. (\ref{Eq:W}) may be computationally demanding. Here, we propose an approximation, valid in the case where the function $G (\mathbf{x})$ changes smoothly over the mirror surface, as it is the case in experimentally relevant situations.
This model is also interesting from the practical point of view, since in many cases adaptive optical systems are implemented using spatial light modulators or segmented deformable mirrors, where the modulation surface is partitioned into small pixels.

Suppose we partition the Fourier plane $\Pi_F$ into small squares
(pixels) of side d. Let's define the rectangular function:
\begin{equation}
\Pi (\mathbf{x}) =
\bigg \{
\begin{array}{cc}
0  & \mbox{if $|\mathbf{x}|>\frac{1}{2}$} \\
1 &  \mbox{if $|\mathbf{x}|< \frac{1}{2}$}
\end{array}
\end{equation}
\\

The pixel (l, m) is identified by:
 \begin {equation}
 \label {num_sigma}
 \sigma_{l, m} (x, y)= \Pi \left[ \frac{x}{d}+l \right] \Pi \left[ \frac{y}{d}+ m \right]
 \end {equation}
 selecting the area:
 \begin {eqnarray}
 \label {num_area}
 \nonumber
 \left( l-\frac{1}{2} \right)d < x < \left( l+\frac{1}{2} \right)d \\
 \left( m-\frac{1}{2} \right)d < y < \left( m+\frac{1}{2} \right)d
 \end {eqnarray}

We approximate the value of the phase in each square by the mean value of the actual phase within the square:
 \begin {equation}
 \label {num_phi}
 \varphi_{lm} = \frac {1}{d^2} \int dx dy \varphi (x, y) \Pi \left[ \frac{x}{d}+l \right] \Pi \left[ \frac{y}{d}+ m \right]
 \end {equation}
 that is:
 \begin {equation}
 \label {num_approxExp}
 e^{i\varphi (x, y)} \approx \sum_{l, m} e^{i \varphi_{l, m} \Pi \left[ \frac{x}{d}+l \right] \Pi \left[ \frac{y}{d}+ m \right]}
 \end {equation}
In this case (see Appendix B for a justification):
 \begin {equation}
 \label {num_identity}
 \sum_{l, m} e^{i \phi_{l, m} \Pi \left[ \frac{x}{d}+l \right] \Pi \left[ \frac{y}{d}+ m \right]} = \sum_{l, m} e^{i \phi_{l, m}} \Pi \left[ \frac{x}{d}+l \right] \Pi \left[ \frac{y}{d}+ m \right]
 \end {equation}
 Substituting this expression in Eq. \ref{Eq:Rtau}, and collecting the integrations one finds:
 \begin {equation}
 R (\tau) \approx \sum_{l, m} \sum_{\lambda, \mu} e^{-i \phi_{lm} - \phi_{\lambda, \mu}} \alpha_{l \lambda} I_{m \mu} (\tau)
 \label {num_approxRt}
 \end {equation}
 where:
 \begin {equation}
 \begin {split}
 \alpha_{l \lambda} &= \int dq_x \int dQ_x \Pi \left[ \frac{f}{kd} q_x - l \right] \Pi \left[ \frac{f}{kd} Q_x - \lambda \right] \\
 & \qquad \qquad \qquad e^{j \frac{2d_1}{k_p} (q_x^2 - Q_x^2) P [q_x + Q_x] P [-(q_x + Q_x)]}
 \end {split}
 \end {equation}

\begin{widetext}
and:
 \begin {equation}
 \begin {split}
 I_{m, \mu} (\tau) &= \int dq_y \int dQ_y \Pi \left[ \frac{f}{kd} q_y - m \right] \Pi \left[ \frac{f}{kd} Q_y - \mu \right] e^{j \frac{2d_1}{k_p} (q_y^2 - Q_y^2)} e^{-j\frac{M}{D}\tau (q_y - Q_y)} \\
 & \qquad  \mbox{Sinc} \left[ ML(q_y + Q_y) \Lambda \left( 1 - \frac{2\tau}{DL} \right)\right] P [q_y + Q_y] P[-(q_y + Q_y)]
 \end {split}
 \end {equation}
 Performing the integrations one gets:
 \begin {equation}
 \label {numAlphaRt}
 \alpha_{l \lambda} = \int dx \tilde{P} (x) \Lambda \left[ \frac{fx}{kd} - (l+\lambda)\right] \mbox{sinc} \left\{ \frac {2dd_1}{f} x \Lambda \left[ \frac{fx}{kd} - (l+\lambda)\right]\right\} e^{i \frac{d d_1}{f} (l-\lambda)x}
 \end {equation}
 and
 \begin {equation}
 \label {num_IRt}
 \begin {split}
 I_{m \mu} (\tau) &= \int dx \tilde{P}(x) \Lambda \left[ \frac{fx}{kd} - (m+\mu)\right] \mbox{sinc} \left\{   MLx \Lambda \left[1 - \frac{2\tau}{DL} \right] \right\}\\
 & \qquad \mbox{sinc} \left\{ \frac {2kd}{f} \left( \frac{2d_1}{k_p}x - \frac{M}{D} \tau\right)  \Lambda \left[ \frac{fx}{kd} - (m+\mu)\right]\right\} e^{i \frac {kd}{f} \left( \frac{2d_1}{k_p}x - \frac{M}{D} \tau\right) (m - \mu)}
 \end {split}
 \end {equation}
\end{widetext}

 A similar expression can be found for the background coincidence rate:
 \begin {equation}
 \label {num_approxR0}
 R_0 \approx \sum_{l, m} \sum_{\lambda, \mu} e^{-i (\phi_{lm} - \phi_{\lambda, \mu})} R^{(x)}_{l \lambda} R^{(y)}_{m \mu}
 \end {equation}
 where:
 \begin {equation}
 \label {numR0x}
 \begin {split}
 R^{(x)}_{l \lambda} &= \int dx \tilde{P} (x) \Lambda \left[ \frac{fx}{kd} - (l-\lambda)\right] \\
 & \qquad \mbox{sinc} \left\{ \frac {2dd_1}{f} x \Lambda \left[ \frac{fx}{kd} - (l-\lambda)\right]\right\} e^{i \frac{d d_1}{f} (l+\lambda)x}
 \end {split}
 \end {equation}
 and
 \begin {equation}
 \label {numR0y}
 \begin {split}
 R^{(y)}_{m \mu} &= \int dx \tilde{P}(x) \Lambda \left[ \frac{fx}{kd} - (m-\mu)\right] \mbox{sinc} \left( MLx  \right)\\
 & \qquad \mbox{sinc} \left\{\frac{2dd_1}{f}x   \Lambda \left[ \frac{fx}{kd} - (m-\mu)\right]\right\} e^{i \left[ \frac{d d_1}{f} (m + \mu) -\frac{ML}{2}\right]x}
 \end {split}
 \end {equation}

The advantage of our numerical approach is that one can calculate
and tabulate the functions $R^{(x)}_{l \lambda}$, $R^{(y)}_{m
\mu}$, $\alpha_{l \lambda}$ and $I_{m \mu} (\tau)$ for a given
configuration, determined by the focal length f, the shape of the
detection apertures, the width of the deformable optics and the
distance between the crystal and the detectors. Then, to calculate
the shape of the interference pattern for a certain phase
distribution on the adaptive optics one just needs to change the
coefficient of a linear combination of the tabulated functions.
This can be a helpful tool to study the effect of specific
aberrations on the temporal interference or to engineer the shape
of the HOM dip.

 In the limit of large detector apertures $\tilde{P} (x) \approx \delta (x)$:
 \begin {equation}
 \alpha_{l \lambda} = \delta (l + \lambda)
 \end{equation}

 \begin{equation}
 I_{m \mu} = \mbox{sinc} \left[  \frac{2kdM}{fD} \tau \right] e^{-i \frac{2kdM}{fD}m \tau} \delta (m+\mu)
 \end {equation}

 So:
 \begin {equation}
 R_{BIG} (\tau) \approx \sum_{l, m} e^{-i (\varphi_{lm} - \varphi_{-l, -m})} \alpha_{l, -l} I_{m, -m} (\tau)
 \end {equation}
 We have a discrete formulation of the even-order spatial phase cancellation effect. What affects the shape of the dip is the difference between the phase of the pixel (l, m) and the phase of the pixel symmetric to it with respect to the origin (-l, -m). Therefore, since for even functions  $\phi_{lm} = \phi_{-l, -m}$, the phase difference is zero and they do not contribute to the coincidentce pattern. This way our approximate solution technique is consistent with the general theory.

\section{Discussion}

\subsection {Examples}
\begin{figure} [ht]
\centering
\includegraphics [width = 7 cm] {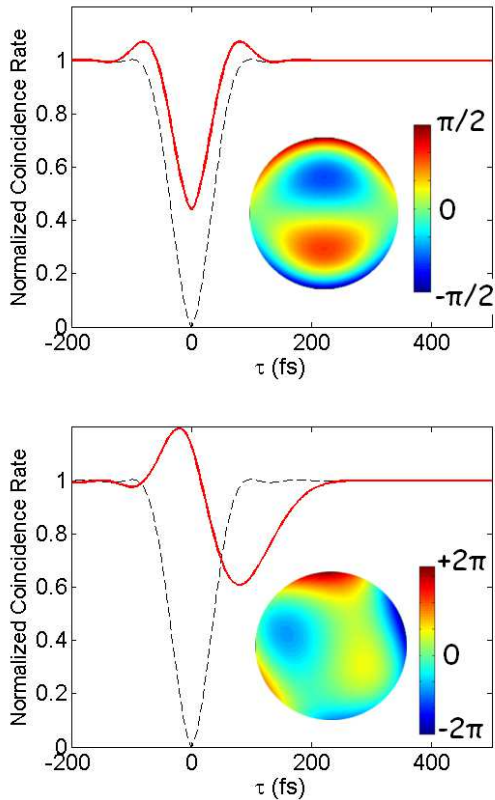}
\caption{Examples of the shape of the polarization
quantum-interference dip with two different spatial phase
modulations (in black the unperturbed dip, in red the modulated
one). In the upper figure the effect of a small amount of coma
along the vertical axis is shown. In the lower figure a more
complicated superposition of aberrations affects dramatically the
shape of the dip} \label {Fig:dips}
\end{figure}

According to Eq. \ref{Eq:Rtau}, the shape of the temporal quantum interference pattern is affected by the spatial modulation given by the deformable mirror. In Section IV we proposed a numerical approach to calculate easily the shape of the temporal interferogram. For example, some results are reported in Fig. \ref{Fig:dips}, for coma (upper plot) and a superposition of several different aberrations (lower plot). The interference visibility clearly degrades in presence of wave-front aberrations.

\subsection {Limits and role of the large aperture approximation}
An interesting question is under what experimental conditions the large aperture approximation can be considered to be valid, so as to obtain the even-order aberration cancellation effect.

\begin{figure} [ht]
\centering
\includegraphics [width = 8 cm] {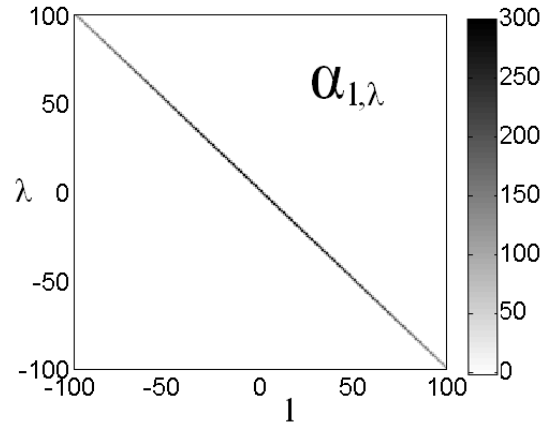}
\caption{Plot of the values of $|\alpha_{l, \lambda}|^2$, for $l,
\lambda = -100...+100$. The radius of the detection apertures is
$R = 5$ mm, the distance between the exit plane of the modulation
section and the detection apertures is $d-1 = 1$ m and the size of
the modulation pixels is $d = 25 \mu$m. Clearly only the diagonal
elements contribute are non-zero, i.e. the ones for which $\lambda
= -l$. In this situation the effect of even-order aberration
cancellation is present.} \label {Fig:alpha}
\end{figure}

\begin{figure} [ht]
\centering
\includegraphics [width = 8 cm] {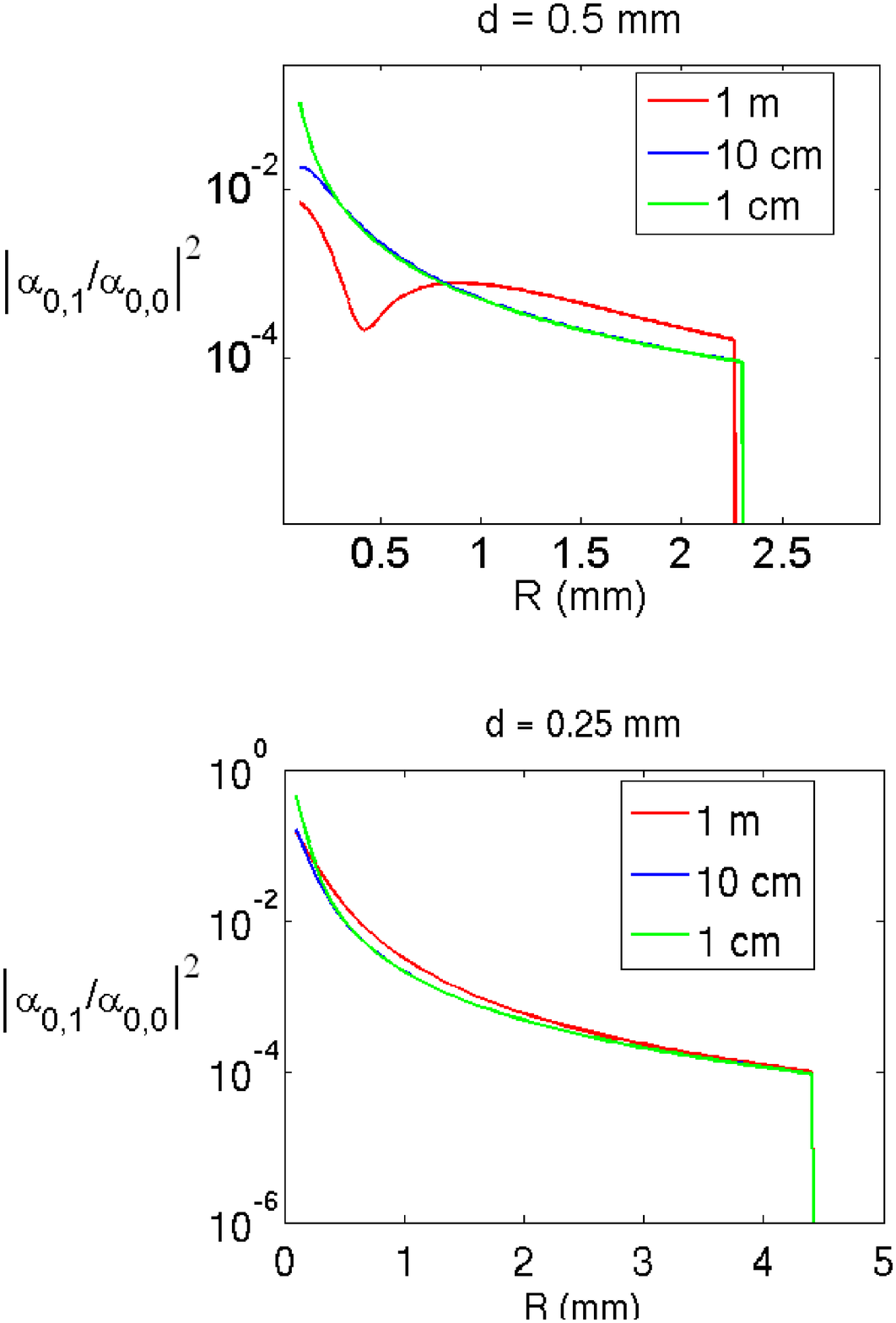}
\caption{Plot of the ratio $\rho_0$ between the intensities of the non-diagonal coefficient $\alpha_{01}$ and the diagonal coefficient $\alpha_{00}$ as a function of the radius of Gaussian detection apertures $R$, for different values of the distance between the exit plane of the modulation section and the detection apertures ($d_1 = 1, 10, 100$ cm). On the upper plot the size of the modulation pixel $d$ is $d = 0.5$ mm, while in the second case it is $d = 0.25$ mm. Clearly, for experimentally interesting cases, the off-diagonal coefficient is at least 3 orders of magnitude smaller than the diagonal one, leading to even-order aberration cancellation.}
\label {Fig:rho}
\end{figure}

According to the numerical approach proposed in Section IV, the
even-order aberration cancellation effect manifests itself in the
limit where, for example:
\begin{equation}
 \alpha_{l, \lambda} \rightarrow \delta (l + \lambda)
\end{equation}

In Fig. \ref{Fig:alpha}, a plot of the value for $\alpha_{l, \lambda}$ is shown for typical values of the relevant experimental parameters (detection aperture radius $R = 5$ mm, detection distance $d_1 = 1$ m and size $d = 0.1$ mm of each pixel in the Fourier plane of the adaptive optical system). Clearly, only the diagonal elements (the ones for which $l = -\lambda$ are significant), suggesting that the effect of even-order aberration cancellation may be observable for most typical experimental parameters.

To get an idea of what happens for different experimental conditions we can compute the ratio between the intensities of the non-diagonal coefficient $\alpha_{01}$ and the diagonal coefficient $\alpha_{00}$:
\begin{equation}
 \rho_0 = \frac{|\alpha_{01}|^2}{|\alpha_{00}|^2}
\end{equation}
The lower the value for $\rho_0$ the less significant the
coefficients for $\lambda \neq -l$ are: the even-order aberration
cancellation effect will therefore manifest itself more clearly.

Values for $\rho_0$ are shown in Fig. \ref{Fig:rho} for two different cases. In both pictures, the value of $\rho_0$ is shown as a function of the Gaussian detection aperture radius $R$,
for three different values of the distance between the plane $\Pi_3$ and the detection lenses $d_1$. In the upper figure, the size of each small square in which the spatial phase is
assumed to be constant is $d = 0.5$ mm, while in the second case it is $d = 0.25$ mm. In both cases $\rho_0$ is significantly smaller than 1, and it becomes smaller and smaller
increasing the value of the detection aperture radius. However, $\rho_0$ is smaller for larger values of $d$, implying that the spatial variability of the modulation phase plays a
role in the degree of even-order cancellation of the modulation itself.\\

It turns out that for the aberration cancellation effect to
appear, it is in fact only necessary for one aperture to be large
and for one detector to be integrated over. This is sufficient to
produce the transverse-momentum delta functions that lead to
even-order cancellation. To demonstrate this, we can for example
consider the case where the aperture at $B$ is large, and the
detector at $B$ is integrated over, while the aperture at $A$ is
taken to be finite, with detector $A$ treated as pointlike. The
location of the pointlike detector will henceforth simply be
denoted as $x$, and we continue to work within the
quasi-monochromatic approximation. If we integrate only over
$x_B$, leaving $x$ unintegrated, then it is straightforward to
show that:
\begin{eqnarray} W^{(0)}& &= \; e^{{icd}\over {\Omega_0}}\\
& & \times\left\{ \tilde Q\left( q+{{\Omega_0x}\over
{cf_0}}\right)\tilde Q^\prime\left( -q^\prime-{{\Omega_0x}\over
{cf_0}}\right)\tilde P_B\left( q^\prime -q\right) \right. \nonumber \\
& &  + \left. \tilde Q\left( -q+{{\Omega_0x}\over
{cf_0}}\right)\tilde Q^\prime\left( q^\prime-{{\Omega_0x}\over
{cf_0}}\right)\tilde P_B\left( q-q^\prime \right) \right\} \nonumber \\
W_M & & = \; e^{{icd}\over {\Omega_0}}\\
& & \times\left\{ \tilde Q\left( q+{{\Omega_0x}\over
{cf_0}}\right)\tilde Q^\prime\left( -q^\prime+{{\Omega_0x}\over
{cf_0}}\right)\tilde P_B\left( -q^\prime +q\right) \right. \nonumber\\
& & +\left. \tilde Q\left( -q+{{\Omega_0x}\over
{cf_0}}\right)\tilde Q^\prime\left( +q^\prime+{{\Omega_0x}\over
{cf_0}}\right)\tilde P_B\left( q-q^\prime \right) \right\}
\nonumber
\end{eqnarray} Here, we have defined $\tilde Q$ and $\tilde
Q^\prime$ to be the Fourier transforms respectively of $p_A$ and
$p_A^\ast$. $\tilde P_B$ is, as before, the Fourier transform of
$|p_B|^2$. We now let aperture $B$ become large, so that the
function $\tilde P_B$ goes over to a delta function. For
$G_1(x)=e^{i\phi(x)}$ and $G_2(x)=1$, we can substitute these
results into the coincidence rate (which will now be a function of
both $\tau$ and the position $x$ of detector $A$), and carry out
the $q^\prime$ and $\nu$ integrals. For the modulation term, we
find: \begin{eqnarray} R_M(x,\tau )&=& R(x,\tau )-R_0(x)\\ &=&
\int d q \; e^{i[\phi(q)-\phi(-q)]}\\ & & \times
e^{{{2iM\tau}\over D}{\mathbf{e_2}\cdot \mathbf{q}}}
e^{{{2iq^2}\over {k_p}}\left[ {{2\tau}\over
D}+L\right]}\mbox{Sinc}\left( {{2q^2L}\over {k_p}}\right)
\nonumber \\ & & \times \left[ \tilde
Q\left(q+{{{\Omega_0x}\over{cf_0}}}\right)\tilde
Q^\ast\left(-q-{{{\Omega_0x}\over{cf_0}}}\right) \right.\nonumber
\\& & +\left. \tilde
Q\left(-q+{{{\Omega_0x}\over{cf_0}}}\right)\tilde Q^\ast\left(
q-{{{\Omega_0x}\over{cf_0}}}\right)\right] .\nonumber
\end{eqnarray} Here, we have used the fact that the Fourier
transform of $p_A^\ast(x)$ equals the complex conjugate of the
Fourier transform of $p_A(x)$, in order to write $\tilde Q^\prime$
in terms of $\tilde Q$. We see from the presence of the factor
$e^{i[\phi(q)-\phi(-q)]}$ that even-order aberration cancellation
occurs even though one aperture is finite and the corresponding
detector is pointlike. This point may be of importance in future
attempts to produce aberration free imaging.

\section{Conclusions}
Summarizing, we have in this paper done a theoretical study of the
relation between the wavelength modulation of the entangled SPDC
photons and the shape of the resulting temporal quantum
interference pattern. Due to the multiparametric nature of the
generated entangled states, the modulation on the spatial degree
of freedom can affect the shape of the polarization-temporal
interference pattern in the coincidence rate. Our aim is twofold:
from one side we want to study the effect of wavefront aberration
on quantum interferometry, from the other we want to discuss a way
to engineer multiparametrically-entangled states.

We have introduced a theoretical model to calculate the shape of
the polarization-temporal interference pattern given a certain
phase modulation in the crystal far-field, assuming as a free
parameter the shape and the dimension of the collection apertures.
Using a numerical method to study the resulting equation has shown
that for typical experimental cases the hypothesis of large
apertures can be assumed valid. In such an approximation, only the
odd part of the assigned phase modulation affects the shape of the
interference pattern. This effect has recently been demonstrated
experimentally \cite{bonatoAberr2008}.

Moreover, it is often useful in experiments to enlarge the
collection aperture in order to collect a higher photon flux and
larger optical bandwidth. But when working with type-II
birefringently phase-matched downconversion, spatial walk-off
between the emitted photons introduces distinguishability between
the two possible events that can lead to coincidence detection,
reducing the visibility of quantum interference. Such walk-off can
be compensated with a linear phase shift in the vertical
direction, restoring high visibility.

\begin{acknowledgments}
This work was supported by a U. S. Army Research Office (ARO)
Multidisciplinary University Research Initiative (MURI) Grant; by
the Bernard M. Gordon Center for Subsurface Sensing and Imaging
Systems (CenSSIS), an NSF Engineering Research Center; by the
Intelligence Advanced Research Projects Activity (IARPA) and ARO
through Grant No. W911NF-07-1-0629 and by the strategic project QUINTET
 of the Department of Information Engineering of the University of Padova. C. B. also acknowledges
financial support from Fondazione Cassa di Risparmio di Padova e
Rovigo.
\end{acknowledgments}

\appendix
\section{Derivation of Eq. \ref{Eq:Rtau}}

In this Appendix we sketch the major steps for the derivation of
Eq. (\ref{Eq:Rtau}). Substituting Eq. (\ref{Eq:acca_j}) into Eq.
(\ref{Eq:prob_ampl2}), and the result into Eq. (\ref{R_tau1}) one
finds the following expressions for $R_0$ and $W_G(\tau)$:

\begin {equation}
\begin{split}
R_0 &= \int d\mathbf{q} d\mathbf{q'} d\nu \Phi^* (\mathbf{q}, \nu)\Phi (\mathbf{q^\prime}, \nu) G_1^* \left( \frac {f}{k} \mathbf{q}\right) G_1 \left( \frac {f}{k} \mathbf{q'}\right)\\
& \qquad \quad \times G_2^* \left( \frac {f}{k} \mathbf{q}\right)
G_2 \left( \frac {f}{k} \mathbf{q'}\right) W^{(0)} (\mathbf {q},
\mathbf{q'}, \nu)
\end{split}
\end {equation}

\begin {equation}
\begin{split}
W_M (\tau) &= \frac{1}{R_0} \int d\mathbf{q} d\mathbf{q'} d\nu \Phi^* (\mathbf{q}, \nu)\Phi (\mathbf{q^\prime}, -\nu) G_1^* \left( \frac {f}{k} \mathbf{q}\right) \\
&  \times G_1 \left( \frac {f}{k} \mathbf{q'}\right) G_2^* \left(
\frac {f}{k} \mathbf{q}\right) G_2 \left( \frac {f}{k}
\mathbf{q'}\right) W (\mathbf {q}, \mathbf{q'}, \nu)
\end{split}
\end {equation}
where:
\begin{widetext}
\begin {equation}
\begin {split}
W^{(0)} (\mathbf{q}, \mathbf{q'}, \nu) &= \int d\mathbf{x_A} d\mathbf{x_B} H^* (\mathbf{x_A}, \mathbf{q}, \nu) H^* (\mathbf{x_B}, -\mathbf{q}, -\nu) H (\mathbf{x_A}, \mathbf{q'}, \nu) H (\mathbf{x_B}, -\mathbf{q'}, -\nu)\\
& \qquad \qquad +  H^* (\mathbf{x_A}, -\mathbf{q}, -\nu) H^* (\mathbf{x_B}, \mathbf{q}, \nu) H (\mathbf{x_A}, -\mathbf{q'}, -\nu) H (\mathbf{x_B}, \mathbf{q'}, \nu)
\end{split}
\end {equation}
and
 \begin {equation}
 \begin {split}
 W (\mathbf{q}, \mathbf{q'}, \nu) &= \int d\mathbf{x_A} d\mathbf{x_B} H^* (\mathbf{x_A}, \mathbf{q}, \nu) H^* (\mathbf{x_B}, -\mathbf{q}, -\nu) H (\mathbf{x_A}, -\mathbf{q'}, \nu) H (\mathbf{x_B}, \mathbf{q'}, -\nu)\\
 & \qquad \qquad +H^* (\mathbf{x_A}, -\mathbf{q}, -\nu) H^* (\mathbf{x_B}, \mathbf{q}, \nu) H (\mathbf{x_A}, \mathbf{q'}, -\nu) H (\mathbf{x_B}, -\mathbf{q'}, \nu)
 \end{split}
 \end {equation}
\end{widetext}
 The angular and spectral emission function $\Phi (\mathbf{q}, \nu)$ is given by:
 \begin {equation}
 \Phi (\mathbf{q}, \nu) = \int dz \quad \Pi \left[ \frac{z}{L}+\frac{1}{2} \right] e^{-i \Delta (\mathbf{q}, \nu)z}
 \label {phieq}\end {equation}
 Performing the integrals over the spatial coordinates $d\mathbf{x_A}$ and $\mathbf{x_B}$ one gets:
 \begin {equation}
 \begin {split}
 W^{(0)} (\mathbf{q}, \mathbf{q'}, \nu) &= e^{i \frac{2 d_1}{k_p} [|\mathbf{q}|^2 - |\mathbf{q'}|^2]} \tilde{P}_A [(\mathbf{q} - \mathbf{q'})] \tilde{P}_B [-(\mathbf{q} - \mathbf{q'})] \\
 & \qquad \qquad + \tilde{P}_A [-(\mathbf{q} - \mathbf{q'})] \tilde{P}_B [(\mathbf{q} - \mathbf{q'})]
 \end {split}
 \end {equation}
and
 \begin {equation}
 \begin {split}
 W (\mathbf{q}, \mathbf{q'}, \nu) &= e^{i \frac{2 d_1}{k_p} [|\mathbf{q}|^2 - |\mathbf{q'}|^2]}  \tilde{P}_A [(\mathbf{q} + \mathbf{q'})] \tilde{P}_B [-(\mathbf{q} + \mathbf{q'})] +\\
 & \qquad \qquad \tilde{P}_A [-(\mathbf{q} + \mathbf{q'})] \tilde{P}_B [(\mathbf{q} + \mathbf{q'})]
 \end {split}
 \end {equation}

Finally, use of the integral
representation for the sinc function (Eq. (\ref{phieq})) allows the $\nu$
integration to be carried out, but at the expense of introducing two integrations over
a pair of new parameters (say $z$ and $z^\prime$). Note the relation
\begin {equation}
\Pi [x] \Pi [x - \alpha] =\left\{ \begin{array}{ll} 1& , \mbox{
if
} -1\le\alpha\le 0 , \; -{1\over 2}\le x\le{1\over 2}+\alpha\\
1& , \mbox{ if
} 0\le\alpha\le 1 , \; -{1\over 2}+\alpha\le x\le{1\over 2}\\
0& ,\mbox{ else .}\end{array} \right.
\end {equation}
From this, it follows that
\begin{equation}\int \Pi [x] \Pi [x - \alpha] dx =\Lambda (\alpha
),\end{equation}
where $\Lambda (\alpha)$ is the triangle function.
These facts allow us to carry out the two z-integrations that arise from the sinc function,
leading to the result shown in
Eq. (\ref{Eq:Rtau}).

\section{Justification of Eq. (\ref{num_identity})}
Suppose to have a set $A$, which can be partitioned into a
collection of disjoint subsets $A_k, k = 1,2,...$:
\begin {equation}
\bigcup_k A_k = A \qquad A_k \cap A_l = \phi \quad \mbox{if} \quad k \neq l
\end {equation}
To each set we can associate a characteristic function:
\begin {equation}
\chi_k (x) =
\bigg \{
\begin{array}{c}
1, \qquad x \in A_k \\
0, \qquad x \notin A_k
\end{array}
\end {equation}
such that:
\begin {equation}
\sum_k \chi_k (x) = \chi_A(x), \qquad \chi_k (x) \chi_l (x) =
\delta_{kl} \chi_k (x),
\end {equation}
where $\chi_A$ is the characteristic function for the full set,
\begin{equation}\chi_A(x)=\bigg \{
\begin{array}{c}
1, \qquad x \in A \\
0, \qquad x \notin A
\end{array}
\end {equation} The term $e^{i \phi_k \chi_k (x)}$ assumes the value
$e^{i \phi_k}$ for $\chi_k (x) = 1$ and the value $1$ for $\chi_k
(x) =0$ ($1 - \chi_k (x) = 1$), so:
\begin {equation}
\begin {split}
e^{i\sum_k \phi_k \chi_k (x)} &= \prod_k e^{i \phi_k \chi_k (x)}\\
&= \prod_k \left[ 1\cdot(1-\chi_k (x)) +e^{i\phi_k} \cdot \chi_k (x) \right] \\
&= \prod_k \left[ 1 + (e^{i \phi_k}-1) \chi_k \right]
\end {split}
\end {equation}
If we express the first few terms we get:
\begin {equation}
\begin {split}
\prod_k &\left[ 1 + (e^{i \phi_k}-1) \chi_k \right] = \\
&= \left[ 1 + (e^{i \phi_1}-1) \chi_1 \right]\left[ 1 + (e^{i \phi_2}-1) \chi_2 \right]...\\
&= 1 + (e^{i \phi_1}-1) \chi_1 + (e^{i \phi_2}-1) \chi_2 +... \\
& \qquad + (e^{i \phi_1}-1) (e^{i \phi_2}-1) \chi_1\chi_2 +\\
& + (e^{i \phi_1}-1) (e^{i \phi_3}-1) \chi_1\chi_3 + .... \\
& \qquad + (e^{i \phi_1}-1) (e^{i \phi_1}-1) (e^{i \phi_1}-1) \chi_1 \chi_2 \chi_3\\
& \qquad \quad +(e^{i \phi_1}-1) (e^{i \phi_2}-1) (e^{i \phi_4}-1) \chi_1\chi_2 \chi_4+ ...
\end {split}
\end {equation}
So that in the end:
\begin {equation}
\begin {split}
e^{i\sum_k \phi_k \chi_k (x)} &= 1+\sum_k \left[  \left( e^{i \phi_k} -1 \right) \chi_k  \right] \\
&= 1+\sum_k e^{i \phi_k} \chi_k  - \sum_k \chi_k \\
&= \sum_k e^{i \phi_k} \chi_k
\end {split}
\end {equation}
Since the square sets we have used in section IV satisfy Eq. (B1),
then the result expressed in Eq. (B5) is valid for our case.


\begin{thebibliography}{13}
\expandafter\ifx\csname natexlab\endcsname\relax\def\natexlab#1{#1}\fi
\expandafter\ifx\csname bibnamefont\endcsname\relax
  \def\bibnamefont#1{#1}\fi
\expandafter\ifx\csname bibfnamefont\endcsname\relax
  \def\bibfnamefont#1{#1}\fi
\expandafter\ifx\csname citenamefont\endcsname\relax
  \def\citenamefont#1{#1}\fi
\expandafter\ifx\csname url\endcsname\relax
  \def\url#1{\texttt{#1}}\fi
\expandafter\ifx\csname urlprefix\endcsname\relax\def\urlprefix{URL }\fi
\providecommand{\bibinfo}[2]{#2}
\providecommand{\eprint}[2][]{\url{#2}}

\bibitem[{\citenamefont{Schroedinger}(1935)}]{schrod35a}
\bibinfo{author}{\bibfnamefont{E.}~\bibnamefont{Schroedinger}},
  \bibinfo{journal}{Naturwissenschaften} \textbf{\bibinfo{volume}{23}},
  \bibinfo{pages}{807} (\bibinfo{year}{1935}).

\bibitem[{\citenamefont{Klyshko}(1967)}]{klyshko67}
\bibinfo{author}{\bibfnamefont{D.~N.} \bibnamefont{Klyshko}},
  \bibinfo{journal}{JETP Letters} \textbf{\bibinfo{volume}{6}},
  \bibinfo{pages}{23} (\bibinfo{year}{1967}).

\bibitem[{\citenamefont{Harris et~al.}(1967)\citenamefont{Harris, Osham, and
  Byer}}]{harris67}
\bibinfo{author}{\bibfnamefont{S.~E.} \bibnamefont{Harris}},
  \bibinfo{author}{\bibfnamefont{M.~K.} \bibnamefont{Osham}}, \bibnamefont{and}
  \bibinfo{author}{\bibfnamefont{R.~L.} \bibnamefont{Byer}},
  \bibinfo{journal}{Phys. Rev. Lett.} \textbf{\bibinfo{volume}{18}},
  \bibinfo{pages}{732} (\bibinfo{year}{1967}).

\bibitem[{\citenamefont{Giallorenzi and Tang}(1968)}]{giallorenzi68}
\bibinfo{author}{\bibfnamefont{T.~G.} \bibnamefont{Giallorenzi}}
  \bibnamefont{and} \bibinfo{author}{\bibfnamefont{C.~L.} \bibnamefont{Tang}},
  \bibinfo{journal}{Phys. Rev.} \textbf{\bibinfo{volume}{166}},
  \bibinfo{pages}{225} (\bibinfo{year}{1968}).

\bibitem[{\citenamefont{Kleinmann}(1968)}]{kleinman68}
\bibinfo{author}{\bibfnamefont{D.~A.} \bibnamefont{Kleinmann}},
  \bibinfo{journal}{Phys. Rev.} \textbf{\bibinfo{volume}{174}},
  \bibinfo{pages}{1027} (\bibinfo{year}{1968}).

\bibitem[{\citenamefont{Atature et~al.}(2002)\citenamefont{Atature, Giuseppe,
  Shaw, Sergienko, Saleh, and Teich}}]{mete04}
\bibinfo{author}{\bibfnamefont{M.}~\bibnamefont{Atature}},
  \bibinfo{author}{\bibfnamefont{G.~D.} \bibnamefont{Giuseppe}},
  \bibinfo{author}{\bibfnamefont{M.}~\bibnamefont{Shaw}},
  \bibinfo{author}{\bibfnamefont{A.~V.} \bibnamefont{Sergienko}},
  \bibinfo{author}{\bibfnamefont{B.~E.~A.} \bibnamefont{Saleh}},
  \bibnamefont{and} \bibinfo{author}{\bibfnamefont{M.~C.} \bibnamefont{Teich}},
  \bibinfo{journal}{Phys. Rev. A} \textbf{\bibinfo{volume}{66}},
  \bibinfo{pages}{023822} (\bibinfo{year}{2002}).

\bibitem[{\citenamefont{Abouraddy et~al.}(2002)\citenamefont{Abouraddy, Nasr,
  Saleh, Sergienko, and Teich}}]{qoct02}
\bibinfo{author}{\bibfnamefont{A.}~\bibnamefont{Abouraddy}},
  \bibinfo{author}{\bibfnamefont{M.~B.} \bibnamefont{Nasr}},
  \bibinfo{author}{\bibfnamefont{B.~E.~A.} \bibnamefont{Saleh}},
  \bibinfo{author}{\bibfnamefont{A.~V.} \bibnamefont{Sergienko}},
  \bibnamefont{and} \bibinfo{author}{\bibfnamefont{M.~C.} \bibnamefont{Teich}},
  \bibinfo{journal}{Phys. Rev. A} \textbf{\bibinfo{volume}{65}},
  \bibinfo{pages}{053817} (\bibinfo{year}{2002}).

\bibitem[{\citenamefont{Nasr et~al.}(2003)\citenamefont{Nasr, Saleh, Sergienko,
  and Teich}}]{qoct03}
\bibinfo{author}{\bibfnamefont{M.~B.} \bibnamefont{Nasr}},
  \bibinfo{author}{\bibfnamefont{B.~E.~A.} \bibnamefont{Saleh}},
  \bibinfo{author}{\bibfnamefont{A.~V.} \bibnamefont{Sergienko}},
  \bibnamefont{and} \bibinfo{author}{\bibfnamefont{M.~C.} \bibnamefont{Teich}},
  \bibinfo{journal}{Phys. Rev. Lett.} \textbf{\bibinfo{volume}{91}},
  \bibinfo{pages}{083601} (\bibinfo{year}{2003}).

\bibitem[{\citenamefont{Franson}(1995)}]{franson92}
\bibinfo{author}{\bibfnamefont{J.~D.} \bibnamefont{Franson}},
  \bibinfo{journal}{Phys. Rev. A} \textbf{\bibinfo{volume}{45}},
  \bibinfo{pages}{3126} (\bibinfo{year}{1995}).

\bibitem[{\citenamefont{Steinberg et~al.}(1992)\citenamefont{Steinberg, Kwiat,
  and Chiao}}]{steinberg92a}
\bibinfo{author}{\bibfnamefont{A.~M.} \bibnamefont{Steinberg}},
  \bibinfo{author}{\bibfnamefont{P.~G.} \bibnamefont{Kwiat}}, \bibnamefont{and}
  \bibinfo{author}{\bibfnamefont{R.~Y.} \bibnamefont{Chiao}},
  \bibinfo{journal}{Phys. Rev. A} \textbf{\bibinfo{volume}{45}},
  \bibinfo{pages}{6659} (\bibinfo{year}{1992}).

\bibitem[{\citenamefont{Bonato et~al.}(2008)\citenamefont{Bonato, Sergienko,
  Saleh, Bonora, and Villoresi}}]{bonatoAberr2008}
\bibinfo{author}{\bibfnamefont{C.}~\bibnamefont{Bonato}},
  \bibinfo{author}{\bibfnamefont{A.~V.} \bibnamefont{Sergienko}},
  \bibinfo{author}{\bibfnamefont{B.~E.~A.} \bibnamefont{Saleh}},
  \bibinfo{author}{\bibfnamefont{S.}~\bibnamefont{Bonora}}, \bibnamefont{and}
  \bibinfo{author}{\bibfnamefont{P.}~\bibnamefont{Villoresi}}
  (\bibinfo{year}{2008}), \bibinfo{note}{arXiv:0807.2909}.

\bibitem[{\citenamefont{Rubin}(1996)}]{rubin96}
\bibinfo{author}{\bibfnamefont{M.~H.} \bibnamefont{Rubin}},
  \bibinfo{journal}{Phys. Rev. A} \textbf{\bibinfo{volume}{54}},
  \bibinfo{pages}{5349} (\bibinfo{year}{1996}).

\bibitem[{\citenamefont{Goodman}(1996)}]{goodmanFO}
\bibinfo{author}{\bibfnamefont{J.~W.} \bibnamefont{Goodman}},
  \emph{\bibinfo{title}{Introduction to Fourier Optics}}
  (\bibinfo{publisher}{McGraw-Hill}, \bibinfo{year}{1996}),
  \bibinfo{edition}{2nd} ed.

\end{thebibliography}
\end{document}